\documentstyle[art12,12pt,epsf]{article}
\begin{document}

\title{
  \begin{flushright}
  {\normalsize
  SLAC-PUB-8516\\
  July 2000\\\vspace{1cm}
  }
  \end{flushright}
Direct Measurement of $A_{b}$ 
at the $Z^0$ Pole Using a Lepton Tag
\footnote{Work supported by Department of Energy contract DE-AC03-76SF00515.}
}
\author{The SLD Collaboration$^{**}$\\
{\it Stanford Linear Accelerator Center}\\
{\it Stanford University,}\\
{\it Stanford, CA, 94309}}
\date{}
\maketitle

\begin{abstract}
We present a direct measurement of the parity violation parameter $A_b$,
 derived from the left-right forward-backward asymmetry of $b$ quarks
 tagged via leptons from semileptonic decays. The lepton identification 
algorithm combines information from tracking, calorimetry and from the SLD 
Cherenkov Ring Imaging Detector. 
The value of $A_b$ is extracted using a maximum
likelihood fit to the differential cross section for fermion production. 
Vertexing information and decay kinematics have been used to discriminate 
among the different sources of tagged leptons. A new treatment of mixing 
effects and of background contamination has been introduced and a new 
vertexing algorithm has been used in the muon analysis. Based on the 1993-1998 SLD sample of 550K hadronic $Z^0$ decays with highly polarized electron beams, we have measured $A_b$ with a $\sim3\%$ statistical error.
\end{abstract}
\vspace{1.6cm}
\begin{center}
{\sl Submitted to the XXX International Conference on High Energy
Physics \\
(ICHEP 2000), Osaka, Japan, Jul/27-Aug/2 2000.

}
\end{center}
\newpage
  
\section{Introduction}
 
Parity violation in the $Zf\bar{f}$ coupling can be measured via the
observables
$A_f = 2v_fa_f/(v^2_f+a^2_f),$
where $v_f$ and $a_f$ represent the vector and
axial vector couplings to fermion $f$.
   The Born-level differential cross section for the reaction
$e^{+} e^{-} \rightarrow  Z^0 \rightarrow f\bar{f}$ is
\begin{equation}
{d \sigma_f} \, / \, {dz} \propto
(1-A_e P_e) (1+z^2) + 2A_f (A_e - P_e) z \, ,
\end{equation}
where $P_e$ is the longitudinal polarization of the electron beam
($P_e > 0$ for right-handed (R) polarization)
and $z = \cos\theta$ is the direction of the outgoing fermion
relative to the incident electron.
In the presence of $e^-$ beam polarization, it is possible to construct
the left-right forward-backward asymmetry
\begin{equation}
{\tilde{A}}^{f}_{FB}(z) = {{[\sigma_{L}^{f}(z) - \sigma_{L}^{f}(-z)] -
[\sigma_{R}^{f}(z) - \sigma_{R}^{f}(-z)] } \over  {[\sigma_{L}^{f}(z) +
\sigma_{L}^{f}(-z)] + [\sigma_{R}^{f}(z) + \sigma_{R}^{f}(-z)] }} = {{|P_e| A_f \, 2z} \over {(1+z^2)}} \, ,
\end{equation}
for which the dependence on the initial state coupling parameter $A_e$
disappears, allowing a direct measurement of the final state coupling
parameters $A_f$. Thus electron beam polarization
permits a unique measurement of $A_f$, independent of that inferred from the
unpolarized forward-backward asymmetry\cite{EWWG} which measures the
combination $A_eA_f$. In addition, the quantity $A_b$ is largely
independent of propagator effects that modify the effective weak mixing
angle, and so is complementary to other electroweak measurements performed
at the $Z^0$ pole. In particular the Standard Model expectation $A_b~=~0.935$
has only a very slight dependence on the top quark and Higgs boson masses.\\
In this paper we present a direct measurement of $A_b$ based on 
identified leptons from semileptonic $B$ hadron decays. The analysis
is based on the full 1993-1998 SLD data sample of 550,000 $Z^0$ decays
and presents the improvements obtained with the addition of vertexing
information provided by the new vertex detector (VXD3) installed in 1996.
 The measurement complements other direct measurements of $A_b$ performed
at SLD, that use momentum-weighted track charge~\cite{victor}, 
vertex charge~\cite{tom1} and identified 
kaons~\cite{tom2} to determine the sign of the underlying quark in $b\bar b$ events. 
The lepton total and transverse momentum (with respect to the nearest
jet), the mass of the event and some topological decay information 
 are used to classify
each event by deriving probabilities for the decays
\hbox{($Z^0 \rightarrow b\bar{b}, b \rightarrow lepton$),}
\hbox{($Z^0 \rightarrow b\bar{b}, \bar{b} \rightarrow \bar{c}
                            \rightarrow lepton$),}
\hbox{($Z^0 \rightarrow b\bar{b}, b \rightarrow \bar{c}
                            \rightarrow lepton$),}
\hbox{($Z^0 \rightarrow c\bar{c}, \bar{c} \rightarrow lepton$),} and 
($Z^0 \rightarrow background$)~\footnote{leptons from light hadron decays,
photon conversions and misidentified leptons}.
The lepton charge ($Q$) provides quark-antiquark discrimination,
while the jet nearest in direction to the lepton approximates the quark
direction. 
The parameter $A_{b}$ is then extracted by a maximum
likelihood fit of these data to the polarized differential cross section,
taking into account the effects of hard gluon radiation.
Although in this approach the polarized asymmetry (2) is not explicitly
formed, the result for $A_b$ maintain its insensitivity to the initial 
state couplings. 

\section{Data Selection and Lepton Identification}
 
The SLAC Linear Collider and its operation with a polarized
electron beam have been described in detail elsewhere \cite{SLC}.
During the running period from 1993-98, the SLC Large Detector (SLD) recorded
an integrated luminosity of 19.1 $pb^{-1}$ with a luminosity-weighted electron
beam polarization of $ |P_e| = 0.729 \pm 0.004 $ (1997-98) at a mean center of
mass energy of 91.27 GeV.
 
   Charged particle tracks are reconstructed
in the Central Drift Chamber \cite{CDC}
and the CCD-based vertex detector \cite{VXD}, in a 
uniform axial magnetic field of 0.6T. The 
combined momentum resolution in the plane perpendicular to the beam axis is 
$ \delta p_{\bot} / p_{\bot} =
\sqrt{(.01)^2+(.0026~p_{\bot}/GeV)^2 \,}$.
 
   The Liquid Argon Calorimeter (LAC) \cite{LAC}
measures the energies of charged and neutral particles
and is also used for electron identification.
The LAC is segmented into projective towers
with separate electromagnetic and hadronic sections.
In the barrel LAC,
which covers the angular range $ |\cos\theta| < 0.82 $,
the electromagnetic towers have transverse
size ${\rm \sim (36~mrad)^2}$ and are divided longitudinally
into a front section of 6 radiation lengths and a
back section of 15 radiation lengths.
The barrel LAC electromagnetic energy resolution is
$ {{\sigma_E} / {E}} = {{15\%} / {\sqrt{E(GeV)}}} $.
 
   Muon tracking is provided by the
Warm Iron Calorimeter (WIC) \cite{WIC}.
The WIC is 4 interaction lengths thick and surrounds the
$2.8+0.7$ interaction lengths of the LAC and SLD magnet coil.
Sixteen layers of plastic streamer tubes interleaved with
2 inch thick plates of iron absorber provide muon hit resolutions
of 0.4 cm and 2.0 cm in the azimuthal and axial directions respectively.
 
   The Cerenkov Ring Imaging Detector (CRID) \cite{CRID} 
measures the velocities of charged tracks using the angles of
\v{C}erenkov photons emitted in liquid and gaseous radiators.
 Only the gas
information has been included in this analysis, since the liquid covers only
marginally the  interesting momentum region ($p > 2$ GeV/c). 
Electrons are distinguishable
from pions in the region between 2 and 5 GeV and the muon
identification (because of pion rejection) also improves considerably in
this region. Kaon and proton rejection also helps the muon
identification up to momenta of 15 GeV.

Hadronic events are selected by requiring at least 15 GeV of energy
in the LAC and at least six tracks with $ p_{\bot} > 250~{\rm MeV} $.
Approximately 550,000 events were found in the 1993-98 data sample,
with negligible background.
Jets are formed by combining calorimeter energy
clusters according to the JADE algorithm \cite{JADE}
with parameter $y_{cut} = 0.005$.
The jet axis closely approximates the $b$-quark direction in
$Z^0 \rightarrow b\bar{b}$ events, with an angular resolution of
${\rm \sim 30~mrad}$.
An electron or muon tag is used to select semi-leptonic decays. 
Electrons are identified with both LAC and CRID information for tracks
with $p>2$~GeV in the angular range $ |\cos \theta| < 0.72$.
 Calorimeter information is used to build discriminant 
variables which exploit the characteristics  of electromagnetic showers,
including transverse and longitudinal shower development shapes,
and matching of LAC energy and track momentum.   
The CRID information is stored in likelihood functions corresponding to  
each particle type hypothesis \cite{CRIDlike}. 
These quantities are used as input variables to a  
 Neural Network, trained on Monte Carlo tracks \cite{donatella}.
The effficiency (purity) for electron identification is on average 62$\%$ 
(70$\%$) and over 78$\%$ (80$\%$) for electrons with momenta greater than 15 GeV/c.
This estimate includes electrons from photon conversions as signal.
As pion misidentification is the largest contribution to the background,
the simulation has been verified using charged pions from reconstructed 
$K^0_s\rightarrow\pi^+\pi^-$ decays. The fraction of such pions misidentified
as electrons is $(1.23\pm 0.15)\%$, consistent with a MC expectation of 
$(1.36\pm0.07)\%$. Electrons from photon conversions are removed from the 
sample with a 70$\%$ efficiency. 
The remaining photon conversion background is clustered at low
momentum, away from most of the signal region.

Muon identification is performed for tracks with $p>2$~GeV
in the angular range $ |\cos\theta| < 0.70 $, although the muon identificaton 
efficiency falls off rapidly for $ |\cos\theta| > 0.60 $ (in the region
between the barrel and the endcaps).
CDC tracks are extrapolated along with the associated error matrices,
including multiple scattering, and matched with hit patterns in the WIC.
 For $ |\cos\theta| < 0.60 $,
87$\%$ of the simulated muon tracks have successful matching between the CDC and the WIC.
 The second step of the muon
identification exploits the information from the CRID. 
The CRID $k - \mu$
separation variable alone rejects $51\%$ of the remaining $k$ and $p$
(with only $2\%$ loss in the signal), while, for $p<6$~GeV,
the  $\pi - \mu$ separation variable rejects $37\%$ of $\pi$ (with $5\%$
loss in the signal).
Since the CRID
information is intrinsically momentum dependent, different sets of cuts on the 
distributions of the discriminant variables have been
optimized in different momentum regions to achieve best purity and efficiency. 
The purity of the final sample is improved by requiring that the candidate muons
fully penetrate the WIC, and by applying further cuts on the number of hits associated with the tracks,
on the $\chi^2$ of the CDC/WIC matching and on the $\chi^2$ of the fit of the track in the WIC.
   MC studies show that pion punch through background is negligible.
Muons from pion and kaon decays and hadronic showers
are a significant background,
but fall off rapidly with increasing momentum. From a study on a pure 
pions data sample, obtained from kinematically selected $K^{0}_s
\rightarrow \pi^+\pi^-$ decays, 
$\sim .3\%$ of pions, with $p>2$~GeV, are identified as muons.
The muon identification efficiency is over $81\%$ with a purity of $68\%$
($8\%$  
misidentified tracks and $24\%$ muons from light hadron decays)  
for $p>2$~GeV and  $ |\cos\theta| < 0.60 $.  
 
\section{Monte Carlo Simulation}

The likelihood that an identified lepton comes from each of the physics sources 
($b\rightarrow{l}$, $b\rightarrow{c}\rightarrow{l}$, $c\rightarrow{l}$,
background etc.) relies directly on MC simulation. 
$Z^0$ decays are generated by the JETSET 7.4 program \cite{LUND}.
The $B$ hadron decay model was tuned
to reproduce existing data from other experiments.
Semileptonic decays of $B$ mesons are generated according to the
ISGW formalism \cite{ISGW} with a 23$\%$ $D^{**}$ fraction,
while semileptonic decays of $D$ mesons are generated with JETSET 
with the 1994 Particle Data Group branching ratios \cite{pdb}.
Particularly important experimental constraints are provided by
the $B \rightarrow lepton$ and $B \rightarrow D$ momentum spectra
measured by CLEO \cite{CLEO} \cite{ISGW2},
the $D \rightarrow lepton$ momentum spectrum measured by
DELCO \cite{DELCO}, and
the $B \rightarrow hadron$ multiplicities measured by ARGUS \cite{ARGUS}.

   The SLD detector response is simulated in detail
using GEANT \cite{GEANT}
and has been checked extensively against $Z^0$ data.

\section{Vertex Mass Reconstruction} 

Vertex identification is done topologically, by searching for space 
points in 3D where track density functions overlap~\cite{davej}. Each track is 
parametrized by a Gaussian probability density tube with a width 
equal to the uncertainty in the measured track position at the IP.
Points that are characterized by a large overlap of these Gaussian
 probability tubes are considered candidate ({\it seed}) vertices.
By clustering maxima in the density distribution, secondary vertices
are found for the two hemispheres. The efficiency for reconstructing
a vertex in the same hemisphere as the lepton is $\sim 66\%$ (1996-98).
The mass of the secondary vertex is calculated using the tracks attached
to the vertex itself. Each track is assigned the mass of a charged pion 
and the invariant mass of the vertex is thus calculated. This is then corrected
to account for neutral particles by using kinematic information.
By comparing the vertex flight path and the momentum sum of the tracks
associated to the secondary vertex, one calculates a minimum amount of 
missing transverse momentum to be added to the invariant ({\it raw}) mass.
This is done by assuming that the true quark momentum is aligned with the 
flight direction of the vertex. The so-called $P_t$-corrected mass is then 
given by:
\begin{equation}
M_{VTX} = \sqrt{M_{raw}^2+P_t^2}+|P_t|
\end{equation}
We require that the transverse momentum contribution be less than the 
initial mass of the secondary vertex, to ensure that poorly measured
vertices in $uds$ events do not leak into the final sample by adding 
large $P_t$. 

\section{Maximum Likelihood Fit}
 
\begin{figure}[t]
\bigskip
\epsfxsize=2.5in   
\epsfysize=2.0in
\centerline{\epsffile{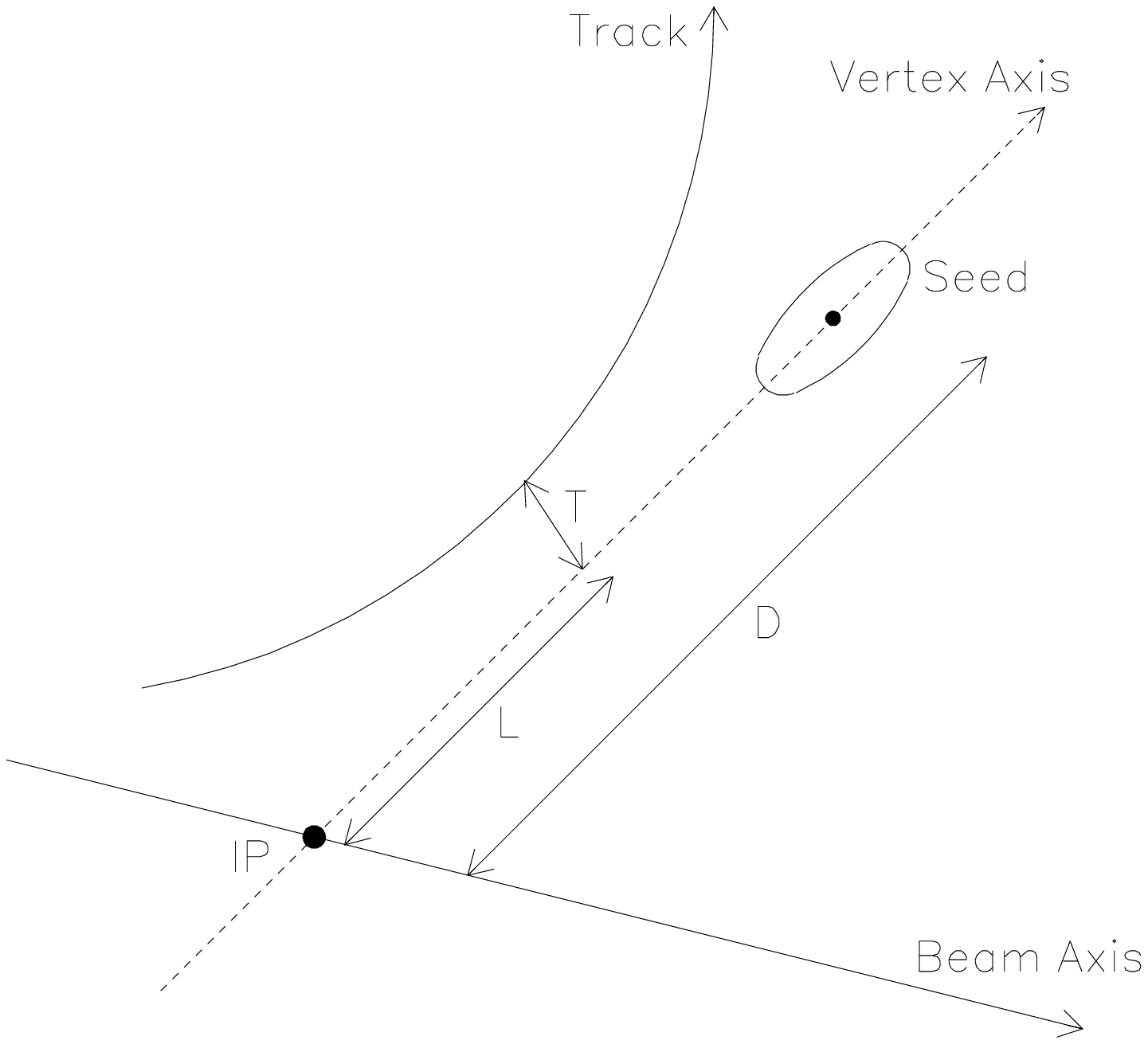}}
\caption{Topological parameters of a track: D is the distance of the secondary seed vertex from the interaction point along the line connecting them; T is the transverse distance of the track from the vertex axis calculated at the point of closest approach (POCA) and finally L is the distance from the IP of the projection of the POCA on the vertex axis.}
\label{topology}
\end{figure}

\begin{figure}
\bigskip
\epsfxsize=2.7in   
\epsfysize=2.7in
\centerline{\epsffile{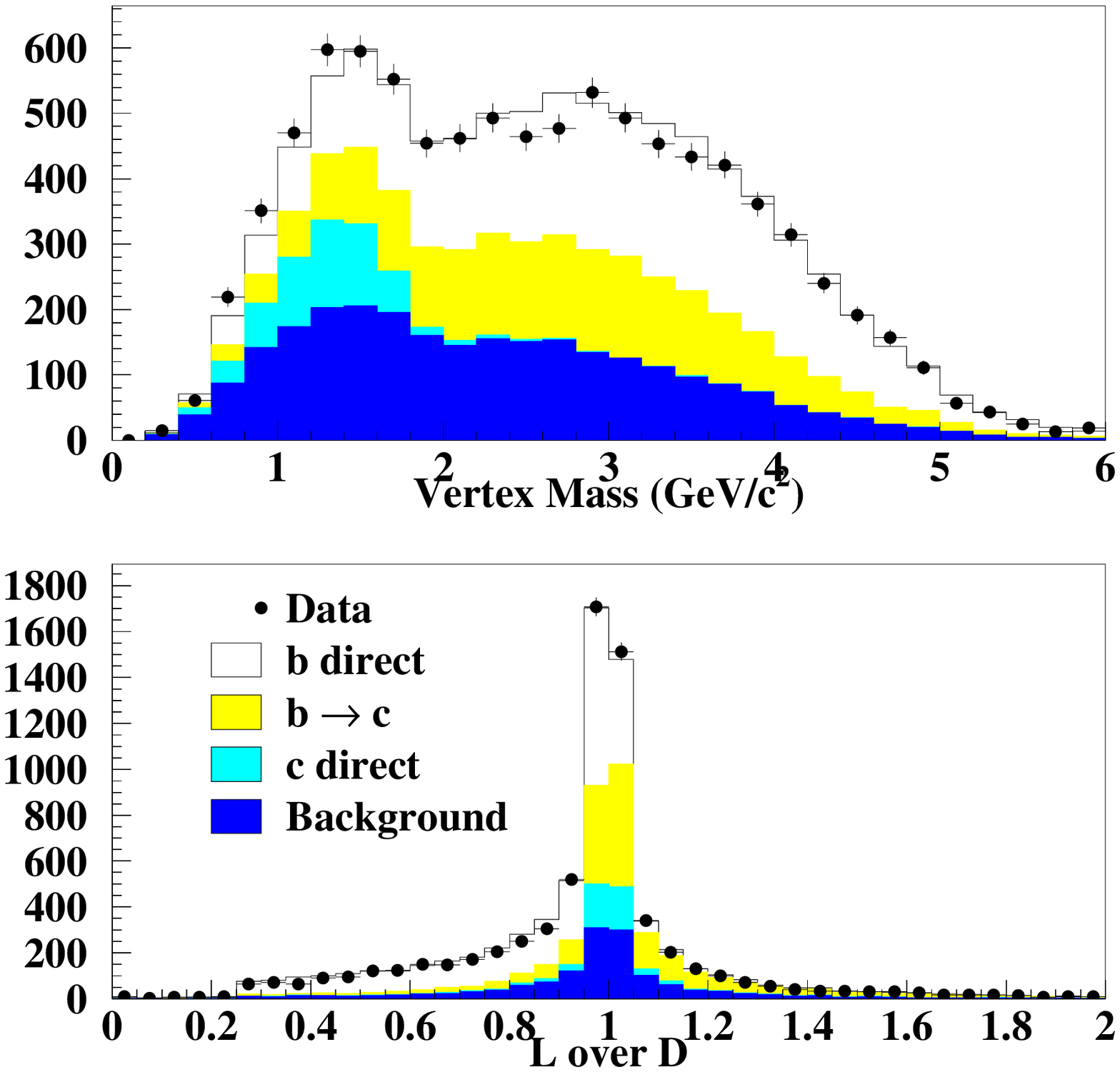}}
\caption{Vertex mass and L/D distributions for electrons in the data (dots) and Monte Carlo (histogram).}
\label{elevtxinfo}
\end{figure}

\begin{figure}
\centering
\parbox{2in} {
\epsfxsize=1.7in   
\epsfysize=1.7in
\epsfbox{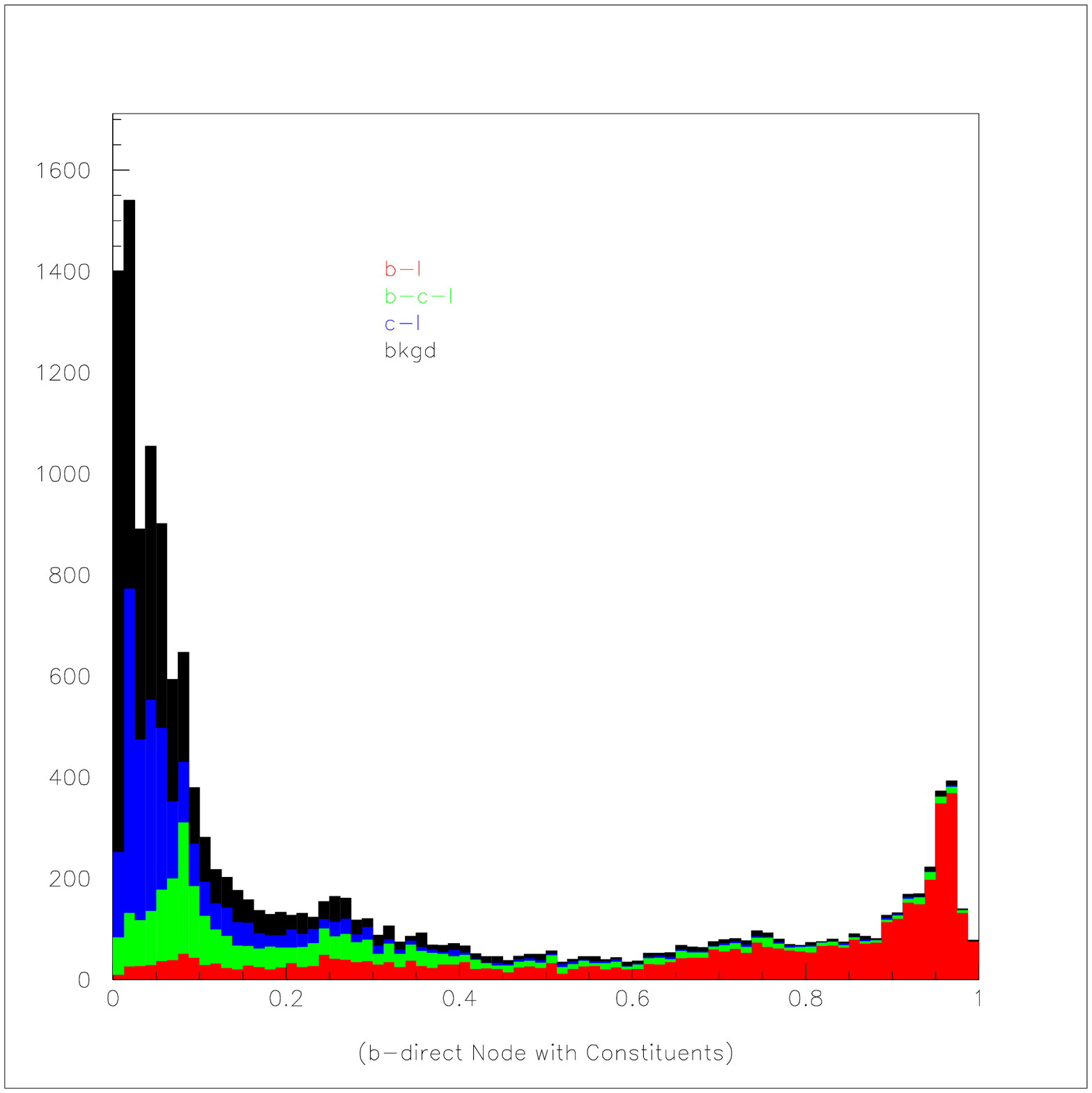}
\caption{Electron NN output, $b\rightarrow{e}$ node.}
\label{nnout1}}\hglue 0.3in\parbox{2.0in}{
\epsfxsize=1.7in   
\epsfysize=1.7in
\epsfbox{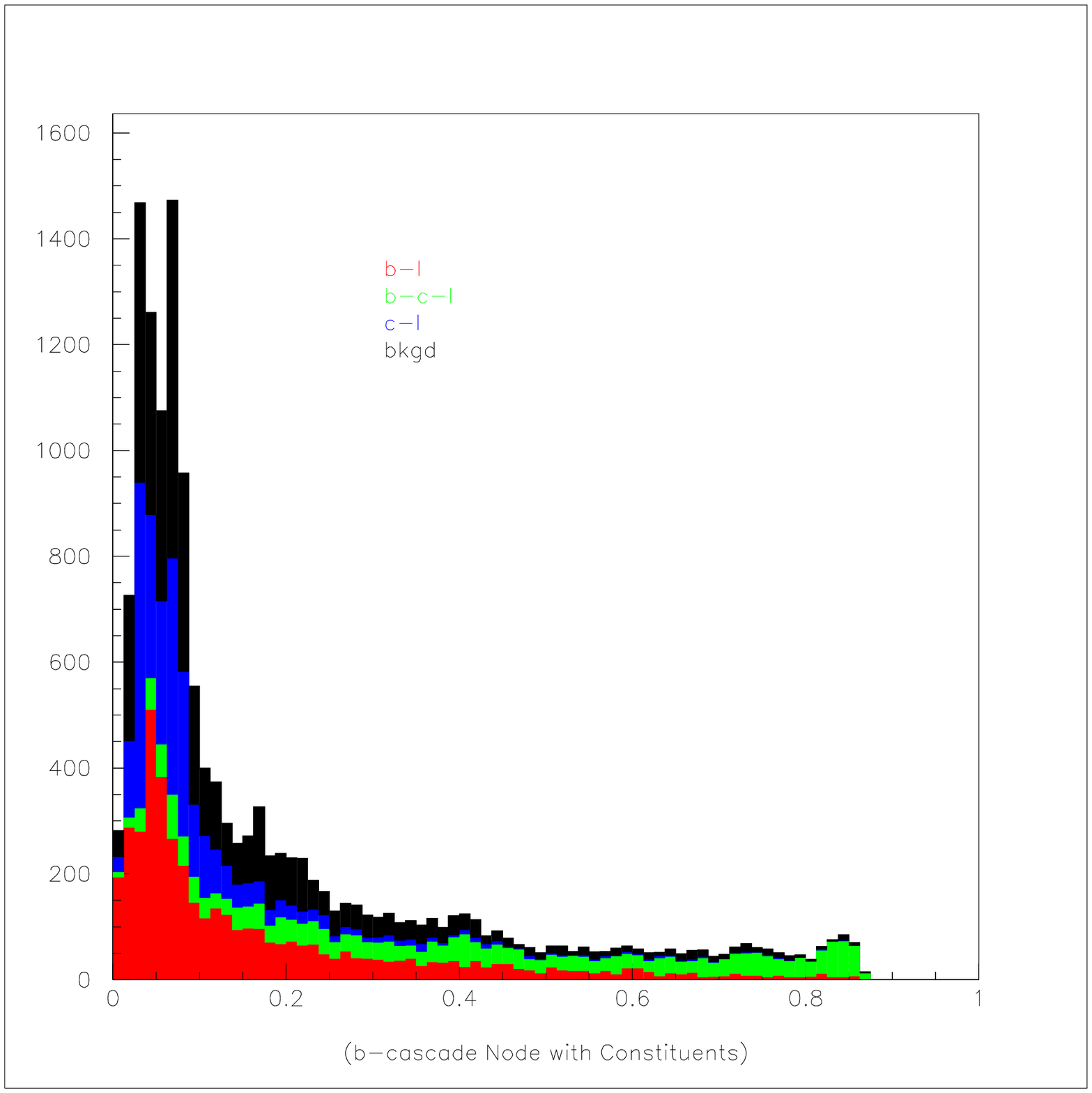}
\caption{Electron NN output, $b\rightarrow{c}\rightarrow{e}$ node.}
\label{nnout2}}
\smallskip
\centering
\parbox{2in} {
\epsfxsize=1.7in   
\epsfysize=1.7in
\epsfbox{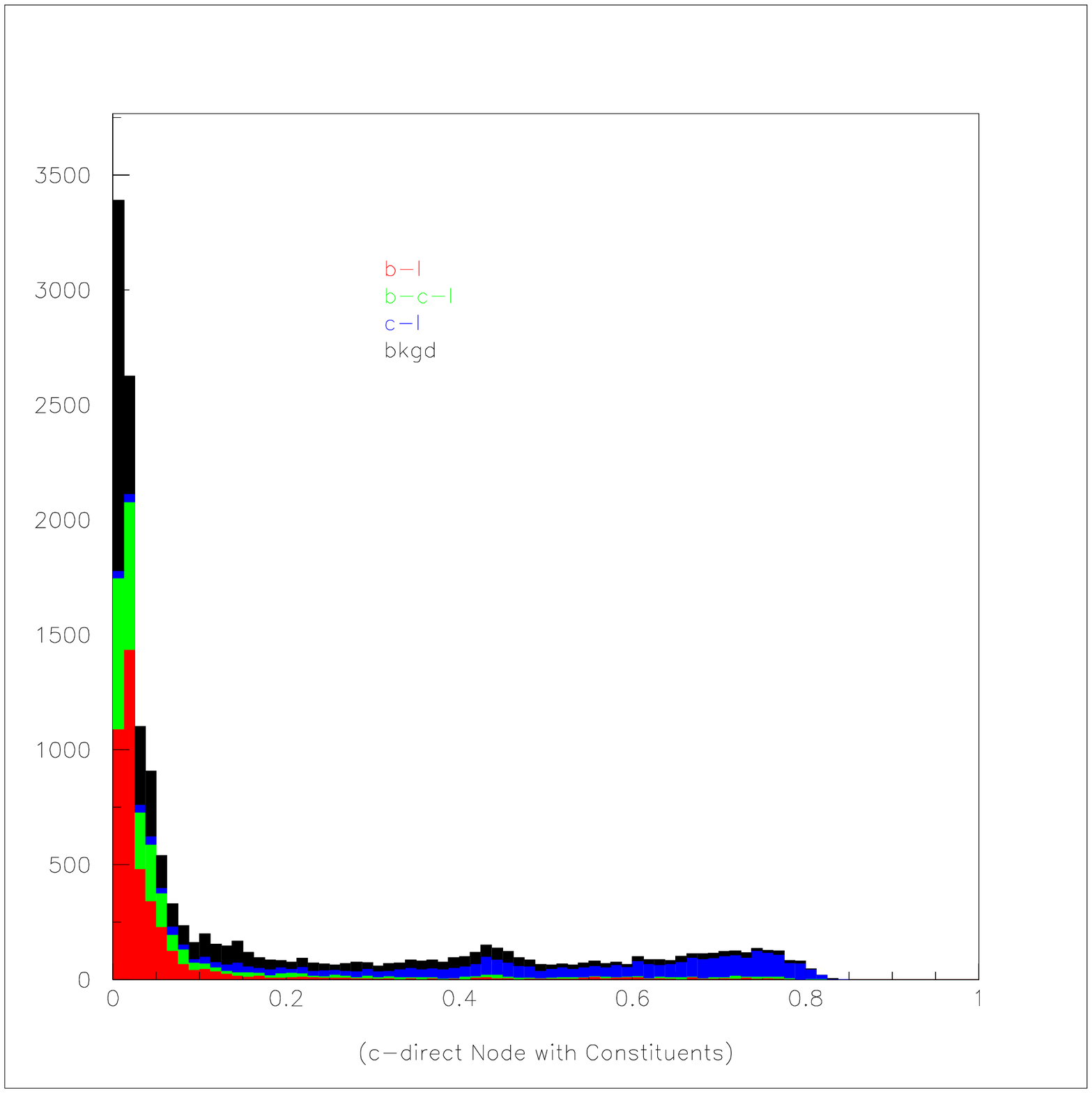}
\caption{Electron NN output, $c\rightarrow{e}$ node.}
\label{nnout3}}\hglue 0.3in\parbox{2.0in}{
\epsfxsize=1.7in   
\epsfysize=1.7in
\epsfbox{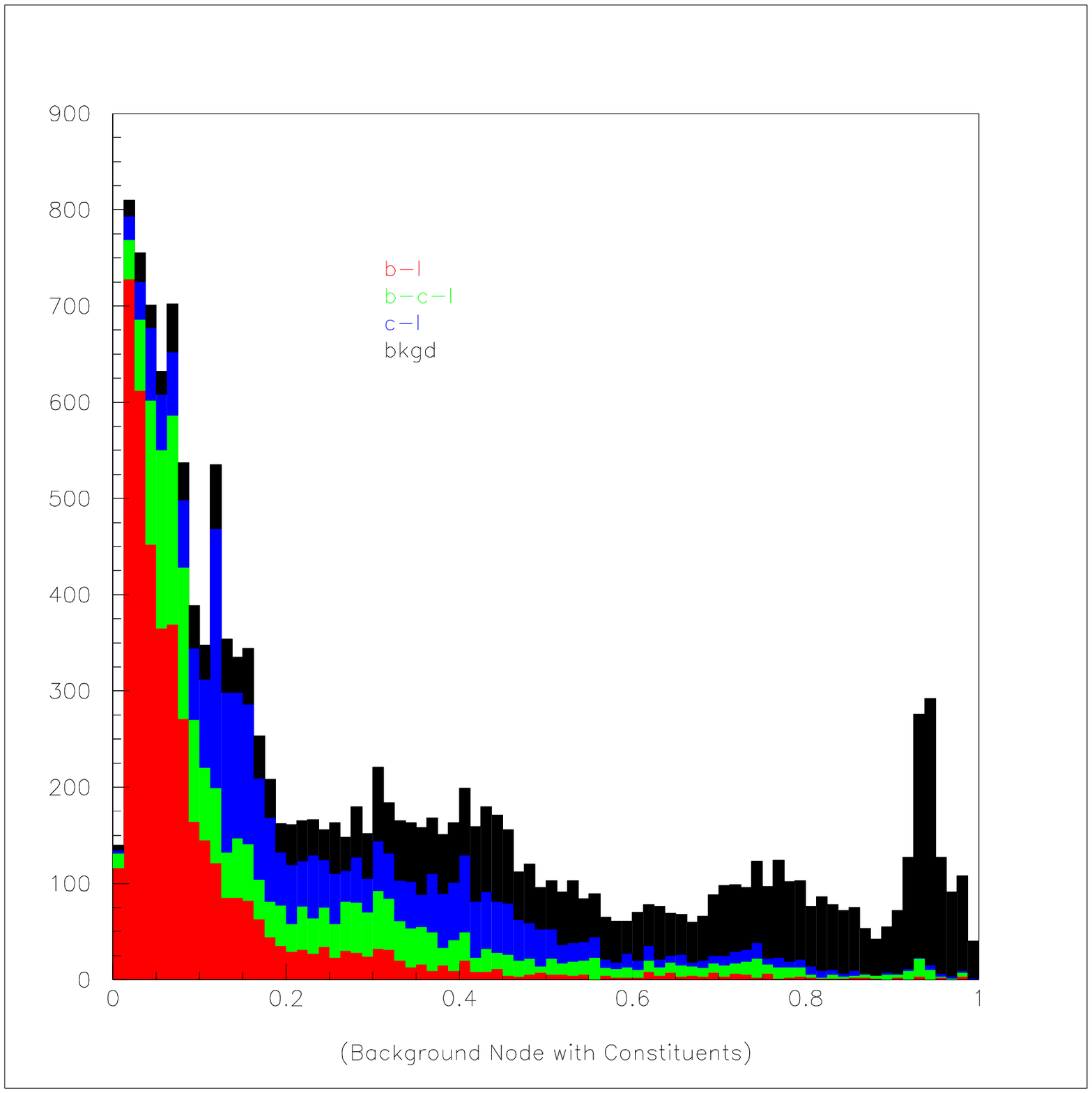}
\caption{Electron NN output, background output.}
\label{nnout4}}
\end{figure}

\begin{figure}
\centering
\epsfxsize=4.0in
\epsfysize= 3.0in 
\centerline{\epsffile{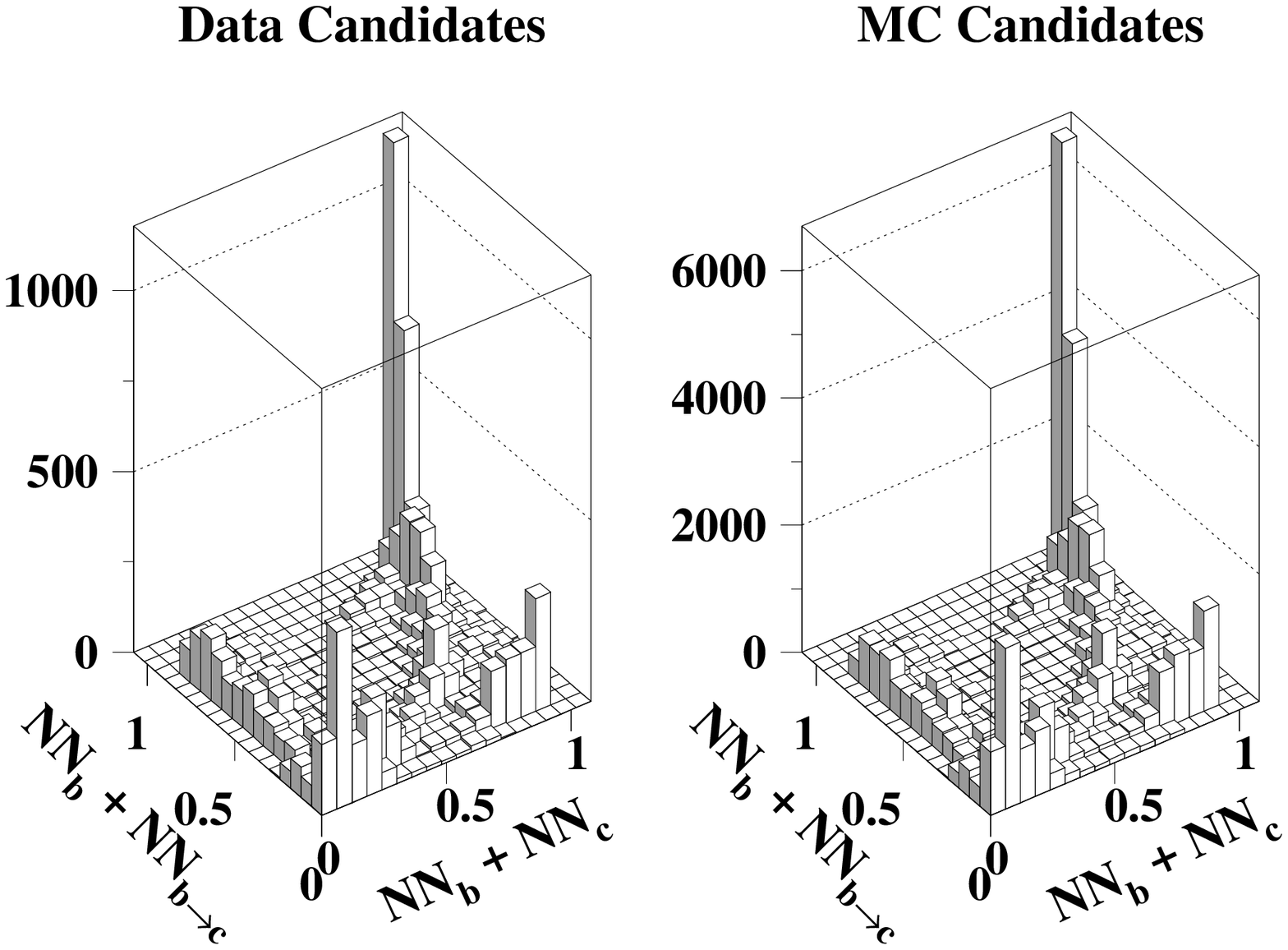}}
\caption{Dalitz space plot of Neural Network output for data and Monte Carlo candidates.}
\label{nndalitz1}
\smallskip
\epsfxsize=4.0in
\epsfysize=4.0in
\centerline{\epsffile{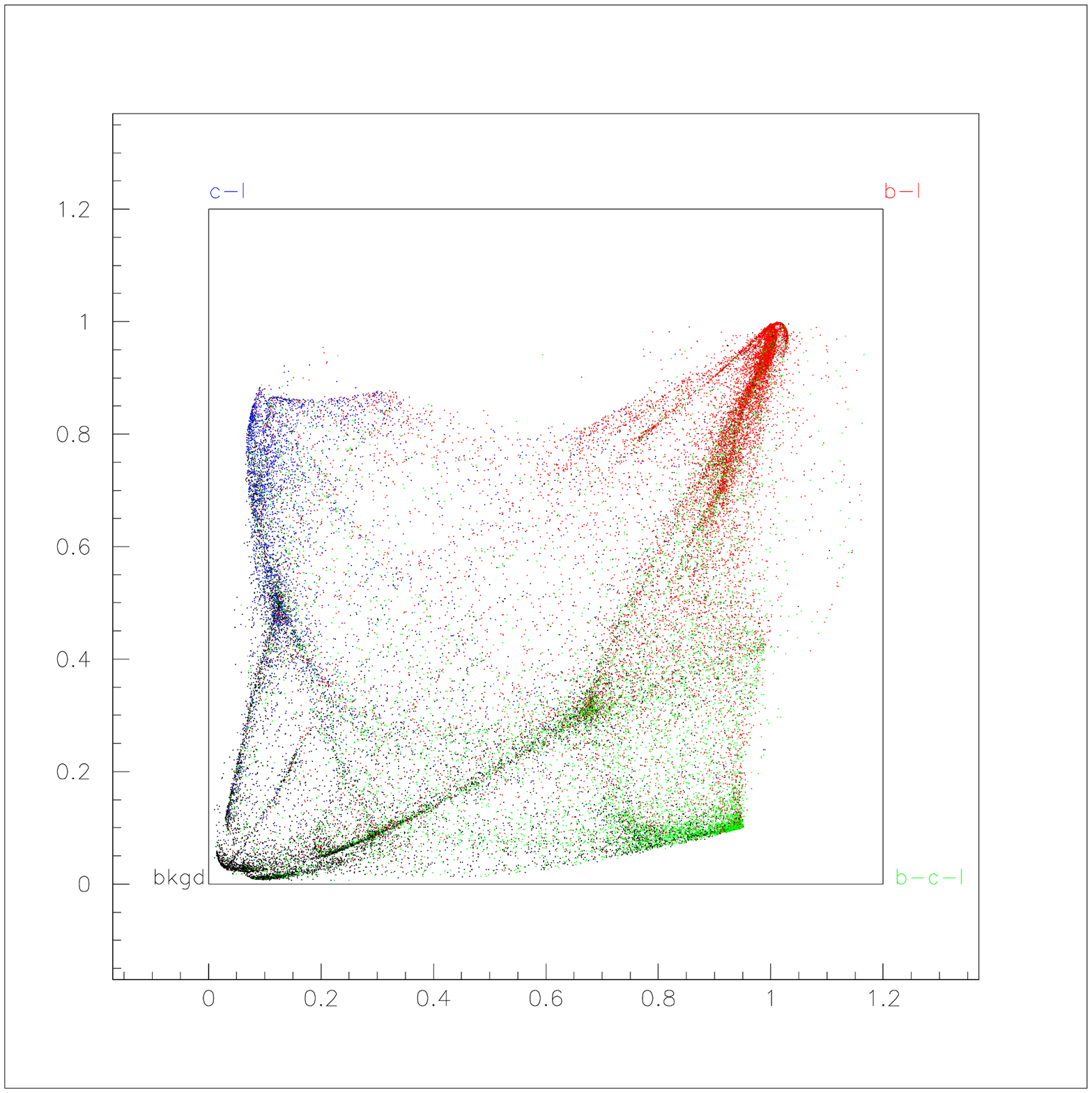}}
\caption{Dalitz space plot for Monte Carlo candidates. Ideally, $b$ direct decays should be clustered around (1,1), $b$ cascade decays at (0,1), charm decays at (0,1) and background events at (0,0).}
\label{nndalitz2}
\end{figure}

Separation between the various lepton sources is accomplished 
using kinematic and vertexing information. Probabilities 
for each of the decay processes are assigned to every lepton, and 
calculated separately for electrons and muons.\\
For electron candidates, eight discriminating variables are
calculated based on characteristics of the event~\cite{jorge}. These
are track momentum ($p$), momentum transverse to the nearest
jet ($p_t$), same hemisphere vertex mass,
same hemisphere vertex momentum, same hemisphere
vertex significance, ratio of the track 
longitudinal distance from the IP along the vertex axis to 
the vertex distance from the IP (L/D, see fig.~\ref{topology}), 
 estimate of the underlying
$b$ quark boost, and the opposite hemisphere vertex mass (fig.~\ref{elevtxinfo}).
(Note: vertexing variables are not always all available for every event, 
but there is a requirement on the  presence of a reconstructed vertex 
in at least one hemisphere.)
These variables are fed into an Artificial Neural Network,
configured with 1 input layer, 2 hidden layers, and 4 output nodes (see figs.~\ref{nnout1},~\ref{nnout2},~\ref{nnout3},~\ref{nnout4}).
The Neural Network weights are set by training on a Monte Carlo
simulation of semileptonic decays of heavy quarks in $Z^0$ decays.
The output of the Neural Network is checked with data. 
The probabilities are assigned by transforming the 3 NN output
nodes onto a 2 dimensional space by $x=NN_{b}+NN_{bc}$ and
$y=NN_{b}+NN_{c}$, illustrated graphically by
figs.~\ref{nndalitz1} and ~\ref{nndalitz2}. This space is divided up into bins and
every candidate is assigned classification probabilites based on
the number of events of each type in its corresponding bin.\\

\begin{figure}
\parbox{3in} {
\epsfxsize=3.0in   
\epsfysize=3.0in
\epsfbox{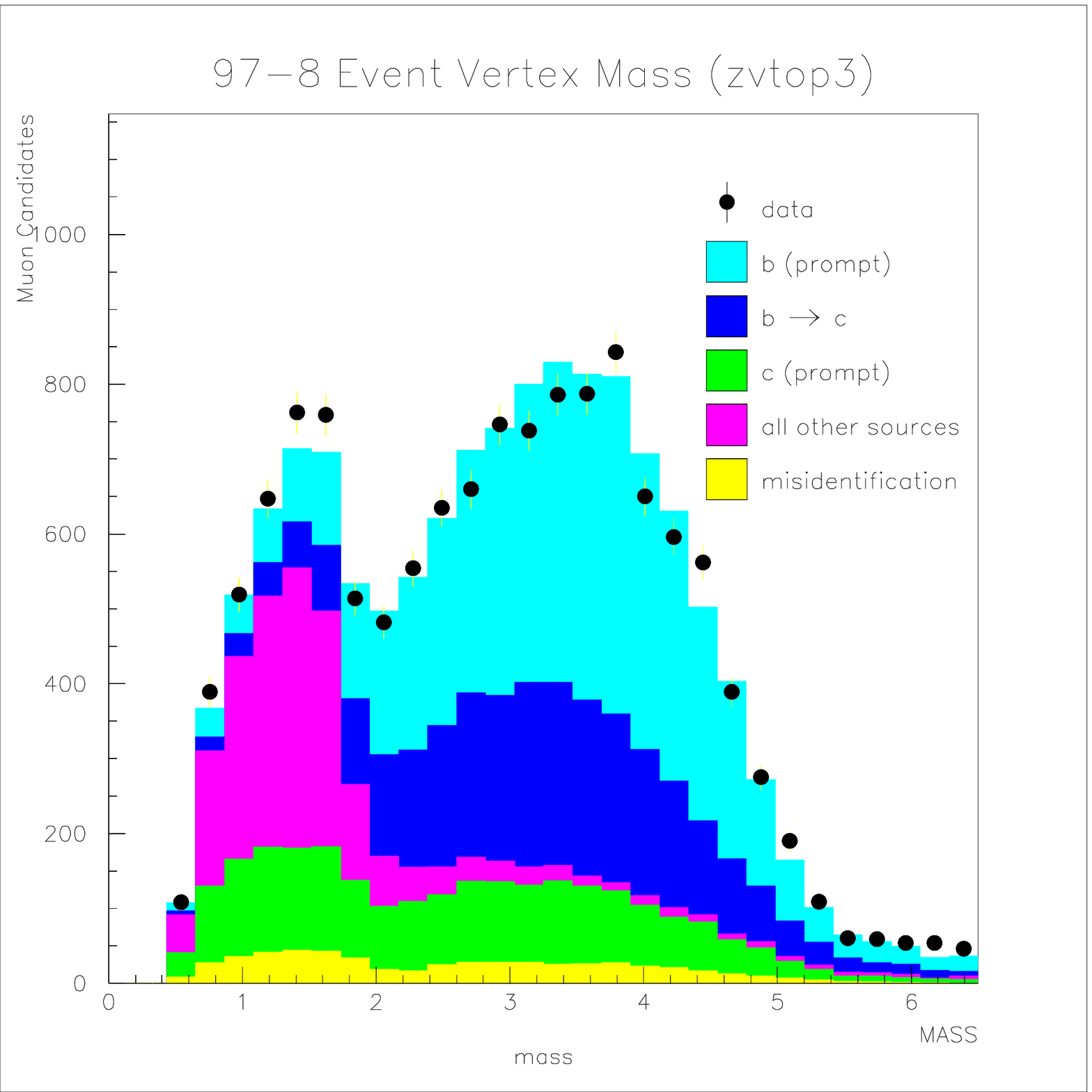}
\caption{Event mass distribution for muons in 1996-8 data (dots) and 
Monte Carlo (histogram). The event mass is assumed to be the highest of the vertex masses found in the two hemispheres.}
\label{muonmass}}\hglue 0.3in\parbox{3.0in}{
\epsfxsize=3.0in   
\epsfysize=3.0in
\epsfbox{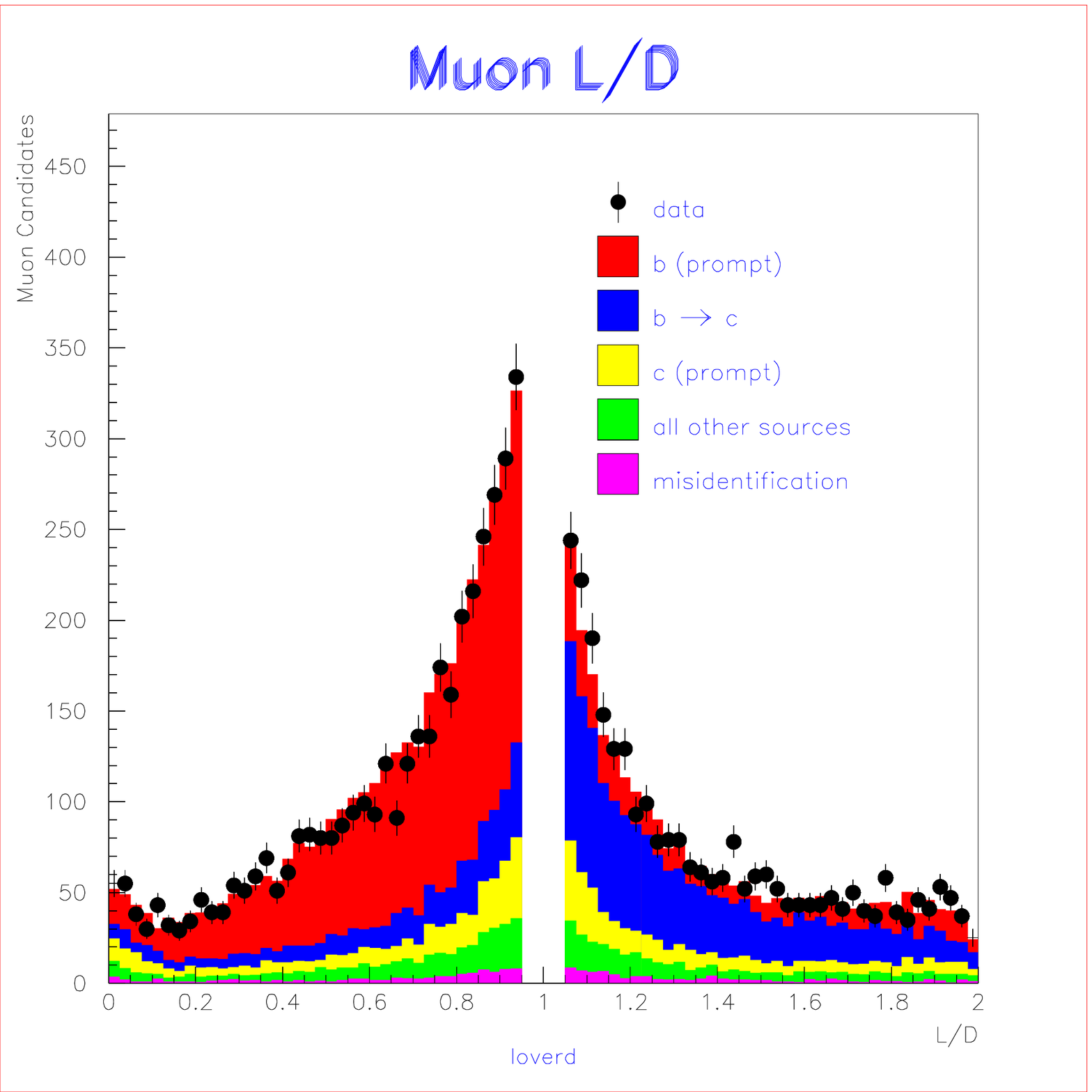}
\caption{L/D distribution for muons in 1996-98 data (dots) and Monte Carlo (histogram).}
\label{ldmuon}}
\end{figure}

\begin{figure}
\bigskip
\epsfxsize=4.5in   
\epsfysize=4.0in
\centerline{\epsffile{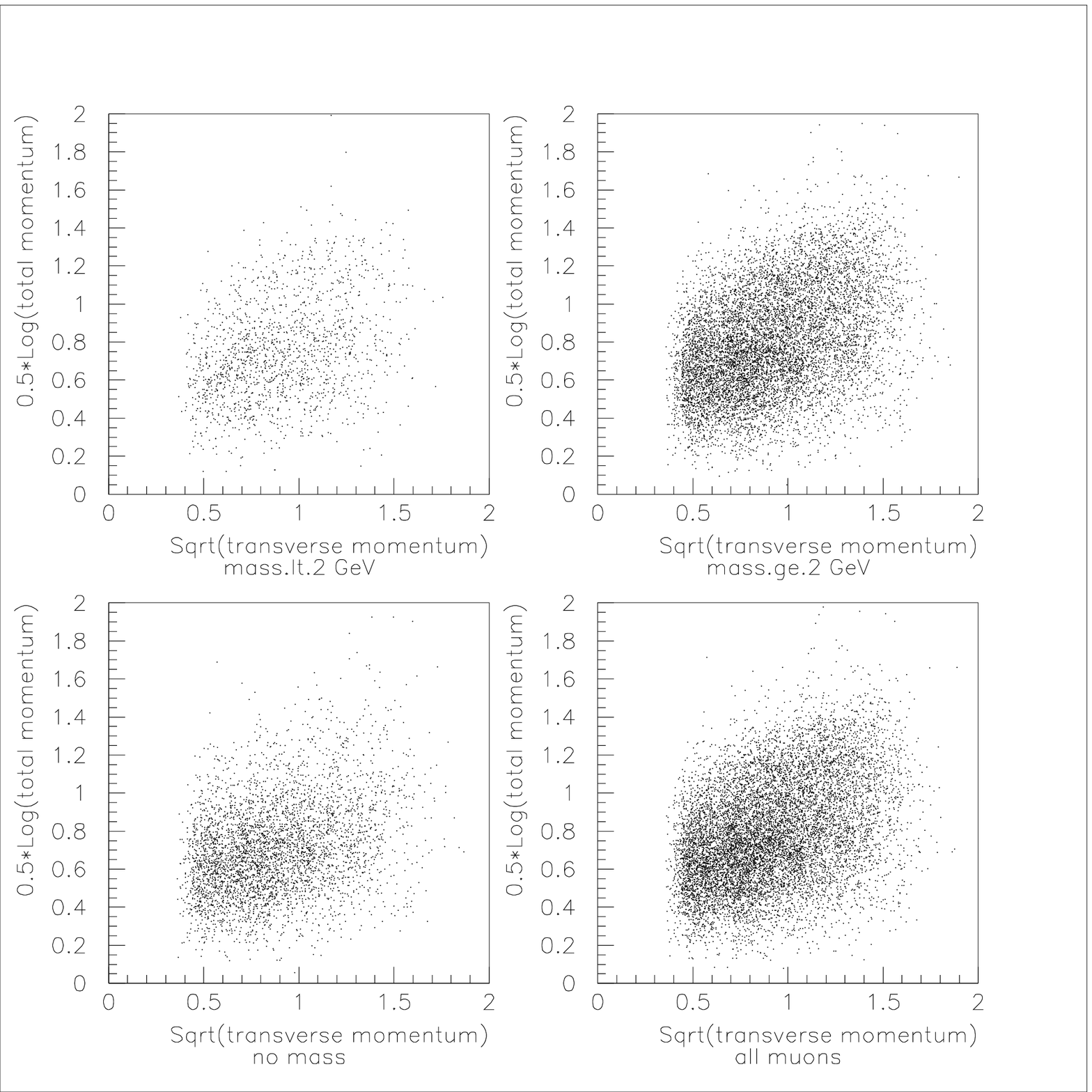}}
\caption{Distribution of $0.5\ln|p|$and $\sqrt{p_T}$ for muons in the 1996-98 data.}
\label{muonMCdistr}
\end{figure}

For the muon candidates, a multi-variate analysis is applied~\cite{us}.
Decay probabilities are calculated for every muon in the data 
by using a {\it nearest neighbours} technique in a 3-D Monte Carlo
phase space. Three planes are defined, corresponding to three different
ranges of the event mass (defined as the largest of the masses 
reconstructed in the two opposite hemispheres of the event, 
fig.~\ref{muonmass}): 
$\ge$ 2 GeV, $2 > M_{vtx} \ge 0.55$ and $0.55 > M_{vtx} \ge 0$ 
(or no vertex found)~\footnote{There is no vertex requirement 
for the muon analysis.}. These planes are parametrized by the 
quantities $\sqrt{P_t}$ and $\ln|P|/2$, to ensure a more 
uniform point distribution (also the scales for the two 
variables are roughly the same, see fig.~\ref{muonMCdistr}). 
The weights for a muon in the data 
are then calculated with a nearest neighbours technique, by selecting
 all MC events within a fixed distance~\footnote{
(optimized with respect to the statistic available)}
 from the data point in the 
corresponding plane and deriving 
the fractions of events of each type in this sample. 
In a second step and for those events only with a reconstructed 
vertex mass, these probabilities are re-weighted with fractions
derived from the MC L/D distribution (see fig.~\ref{ldmuon}), 
which account for the likelihood of an event coming from a certain source
to have a value of L/D included in a  specific interval.
This information helps particularly to enhance the ability 
in separating $b$ direct decays from $b$ cascade decays (which have 
the ``wrong'' charge association).
Correlations between all the different physical quantities employed 
have been taken into account. 
A cross-check via a neural network approach
(similar to the one used for the electrons) has been performed 
and it has given consistent results. \\

A maximum likelihood analysis of all hadronic $Z^0$ events
containing leptons is used to determine $A_{b}$.
The likelihood function contains the following probability term
for each lepton in the data:
\begin{eqnarray}
 P\,(p,p_t,mass,L/D \, P_e,z ; \, A_b,A_c)& & \propto  = \left\{(1+ z^2)(1-A_eP_e)-2Q(A_e-P^i)\right. \nonumber \\
& & \left[(f_b-f_{bc}+f_{b\bar c})(1-2\chi_i)(1-\Delta^b_{QCD}(z))A_b\right. \nonumber \\
& &
\left.\left.+f_c(1-\Delta^c_{QCD}(z))A_c+f_{bkg}A_{bkg}\right]z
\right\}.
\end{eqnarray}
where $z = \cos\theta_{jet}$.
The three signs governing the left-right forward-backward asymmetry ---
beam polarization $P_e$, lepton charge $Q$, and jet direction
$\cos\theta$ --- are incorporated automatically into the
maximum likelihood probability function.
   The fractions ($f_b, f_{bc}, f_{b\bar c}, f_c, f_{bkg}$) are the 
lepton decay probabilities for the different decay modes, derived 
from the MC simulation as described before.
A correction factor $(1-2\chi_i)$ is applied to all b-quark lepton
sources to account for asymmetry dilution due to $B^0\bar{B}^0$ mixing.
For the electrons an average $\bar\chi= 0.1260$ is calculated from Monte Carlo 
(to account for the analysis bias introduced by the vertex requirement),
 and rescaled for $b$ cascade events to account for different mixing
probabilities ($\bar\chi = 0.1364$ and $\bar\chi = 0.1376$ are used 
for $b\rightarrow{c}\rightarrow{e}$ and $b\rightarrow{\bar c}\rightarrow{e}$
events respectively). For the muon analysis, $\chi$ is calculated event 
by event using the truth mixing information of MC events closest to the data
point in a phase space parametrized by $(p, p_t, mass, L/D)$. The dependency 
on the event lifetime is thus automatically accounted for without any 
further need of rescaling. 
The background asymmetry $A_{bkg}$
is derived for the electron analysis as a function of $p$ and $p_t$
from tracks in the data not identified as leptons.
For muons instead, it is calculated as a function of $p$ and $p_t$ 
(using the same
procedure) from MC true background muons, divided in two samples: 
misidentified 
muons and light hadron muons (or misassociated tracks).
A $cos \theta$ dependent QCD correction factor
is applied to the theoretical asymmetry function
to incorporate known QCD corrections to the cross section.
The quantity $\Delta^f_{QCD}(z)$ has been calculated at $O(\alpha_s)$
for massive final state quarks by Stav and Olsen~\cite{stavolsen}
and is as large as 5$\%$
for the b quark at $z\!=\!0$. For an unbiased sample 
of $b\bar b$ events with $|z|<0.7$, correcting for this effect
increases the
asymmetry by $3\%$ overall. However, the theoretical calculations assume perfect
efficiency in the reconstruction of events with emission of gluons
of any energy. The inefficiency of the detector and the presence of cuts
and weights in the analysis cause biases in the event selection which
favor $q \bar q$ events over $q \bar q g$ events, therefore the QCD
correction to be applied is less than the theoretical one. The effects of
this and other related biases have been studied with a MC simulation 
of the analysis chain  and corrected for in
the likelihood function as a function of $\theta$, decreasing the
theoretical QCD correction by about 30$\%$~\cite{OURQCD}.
$O(\alpha_s^2)$ QCD effects are mainly due to two contributions: gluon splitting and second order hard gluon radiation. The gluon splitting correction is calculated apart by re-fitting for $A_b$ (in a MC as data study) after having excluded from the MC all the $g\rightarrow{b\bar b}$, $g\rightarrow{c\bar c}$ events. The difference in the central values, rescaled by the ratio of the current world measurements (OPAL)~\cite{EWWG} of the gluon splitting fractions to the JETSET input values, is assumed as correction. For the second order gluon radiation effects, recent theoretical calculations by Ravindran and van Neerven~\cite{RvNqcd} have been implemented. These have been worked out for different values of the quark masses (pole masses or running) and they predict effects about four times as big as in previous calculations ($\sim 1\%$ correction on $A_b$). 

\section{Results and Systematic Errors} 

The results obtained for the 1993-98 data are as follows,
where the combined result takes into account the systematic
correlations between the muon and electron analyses.
\begin{eqnarray}
Muons & & A_b = 0.950~\pm~0.038_{\it stat}~\pm~0.026_{\it syst} \\  \nonumber
Electrons & &  A_b = 0.876~\pm~0.045_{\it stat}~\pm~0.028_{\it syst} \\ \nonumber
Combined & &  A_b = 0.922~\pm~0.029_{\it stat}~\pm~0.024_{\it syst} \\ \nonumber
\end{eqnarray}

 A list of systematic errors is shown in Table~\ref{systable}.
The background levels have been studied with the MC, but also with a
data sample of pure pions from $K^{0}_s$ decays. The asymmetry of the
background has been varied by $\pm40\%$ of itself for the electron analysis, 
and just rescaled by the ratio of the asymmetry in data and MC for  
charged non-leptonic tracks in the muon case. 
Uncertainty in the jet axis simulation can affect the asymmetry
measurement by distorting the lepton $p_t$ spectrum and, to a lesser
extent, the jet direction. The resulting
systematic error has been studied by comparing the back-to-back direction
of jets for  data and MC in two jet events. The electron sample
is more sensitive to such effects since both jet finding and
electron identification algorithm rely on the same calorimeter response.
The precision of the $ B^{\pm} $ and $ B^0 $ lepton spectra is
directly related to the uncertainty in the $ D^{**} $
branching fraction reported by the CLEO collaboration~\cite{CLEO}.
The systematic error due to uncertainties in the D lepton spectrum has
been estimated by constraining the ACCMM model~\cite{ACCM}
to the DELCO  $D\rightarrow l$ data~\cite{DELCO}.
The systematic error due to the QCD correction includes uncertainties
in the 2nd order QCD calculations for hard gluon emission and gluon
splitting, in the value of $\alpha_s$, and in the
bias due to event selection criteria in the analysis.
This analysis is independent of tracking efficiency, unless such
efficiency depends on $p$, $p_t$ or is not symmetric in $\cos\theta$.
The extent of this $p$ and $p_t$ dependence has been
constrained by reweighting MC tracks by the ratio of the number of tracks in
data and MC as a function of $p$ and $p_t$. The extracted value of $A_f$
is much less sensitive to potential differences in
the relative efficiency for selecting leptons between the forward and
backward hemispheres than are the values of $A_f$ extracted from the
unpolarized forward-backward asymmetry. The relative suppression factor
is greater than $1/{A_e}^2 \sim 50$ for any value of $|z|$ and therefore
forward-backward asymmetry in the detector acceptance is not a
significant source of measurement bias.
$A_c$ has been fixed in the maximum likelihood fit to its 
Standard Model value, and a systematic error has been calculated 
by varying this number by plus or minus twice the current 
statistical uncertainty on the world average of the $A_c$ measurements.\\

The value obtained for $A_b$ from leptons can be combined with 
the other measurements performed at the SLC/SLD, respectively based on
a momentum weighted track charge method, a vertex charge method 
and kaon decays. The resulting SLD average 
\begin{center}
$ A_b = 0.914~\pm~0.024,$
\end{center}
obtained using the data collected in 1993-1998, is consistent with the SM
prediction $A_b=0.935$ and in agreement with recent preliminary
results from LEP and SLD\cite{EWWG}.

\section{Conclusions}
In conclusion, we have measured the extent of parity violation in the
coupling of $Z^0$ bosons to $b$ quarks by using identified charged
leptons from semileptonic decays. The analysis presented in this paper 
is based on the entire sample of 550,000 $Z^0$ decays collected in 1993-98
at SLD and employs vertexing information to separate the different decay 
sources.
The resulting 1993-98 measurement 
\begin{center}
$ A_b =  0.922~\pm~0.029~\pm~0.024,$
\end{center}
represents an improvement
relative to previous measurements\cite{lepton99}.

\section{Acknowledgements}
We thank the staff of the SLAC accelerator department for their
outstanding efforts on our behalf. This work was supported by the
U.S. Department of Energy and National Science Foundation, the UK Particle
Physics and Astronomy Research Council, the Istituto Nazionale di Fisica
Nucleare of Italy and the Japan-US Cooperative Research Project on High
Energy Physics.

\def\ubf#1{{ \bf{#1}}}

%XXXXXXXXXXXXXXXXXXXXXXXXXXXXXXXXXXXXXXXXXXXXXXXXXXXX

%
% author list for inclusion in LaTeX documents
% using \author{} and \address{} commands
%
% Instion number definitions:
%
\section*{$^*$ List of Authors}
\begin{center}
\def\iAOMORI{$^{(1)}$}
\def\iBRI{$^{(2)}$}
\def\iBRUN{$^{(3)}$}
\def\iBU{$^{(4)}$}
\def\iCOLO{$^{(5)}$}
\def\iCSU{$^{(6)}$}
\def\iFERR{$^{(7)}$}
\def\iFRAS{$^{(8)}$}
\def\iJHU{$^{(9)}$}
\def\iLBL{$^{(10)}$}
\def\iMASS{$^{(11)}$}
\def\iMISSI{$^{(12)}$}
\def\iMIT{$^{(13)}$}
\def\iMOSCOW{$^{(14)}$}
\def\iNAGO{$^{(15)}$}
\def\iOREG{$^{(16)}$}
\def\iOXF{$^{(17)}$}
\def\iPERU{$^{(18)}$}
\def\iRAL{$^{(19)}$}
\def\iRUTG{$^{(20)}$}
\def\iSLAC{$^{(21)}$}
\def\iSOONG{$^{(22)}$}
\def\iTENN{$^{(23)}$}
\def\iTOHO{$^{(24)}$}
\def\iUCSB{$^{(25)}$}
\def\iUCSC{$^{(26)}$}
\def\iVAND{$^{(27)}$}
\def\iWASH{$^{(28)}$}
\def\iWISC{$^{(29)}$}
\def\iYALE{$^{(30)}$}

  \baselineskip=.75\baselineskip
\mbox{Koya Abe\unskip,\iTOHO}
\mbox{Kenji Abe\unskip,\iNAGO}
\mbox{T. Abe\unskip,\iSLAC}
\mbox{I. Adam\unskip,\iSLAC}
\mbox{H. Akimoto\unskip,\iSLAC}
\mbox{D. Aston\unskip,\iSLAC}
\mbox{K.G. Baird\unskip,\iMASS}
\mbox{C. Baltay\unskip,\iYALE}
\mbox{H.R. Band\unskip,\iWISC}
\mbox{T.L. Barklow\unskip,\iSLAC}
\mbox{J.M. Bauer\unskip,\iMISSI}
\mbox{G. Bellodi\unskip,\iOXF}
\mbox{R. Berger\unskip,\iSLAC}
\mbox{G. Blaylock\unskip,\iMASS}
\mbox{J.R. Bogart\unskip,\iSLAC}
\mbox{G.R. Bower\unskip,\iSLAC}
\mbox{J.E. Brau\unskip,\iOREG}
\mbox{M. Breidenbach\unskip,\iSLAC}
\mbox{W.M. Bugg\unskip,\iTENN}
\mbox{D. Burke\unskip,\iSLAC}
\mbox{T.H. Burnett\unskip,\iWASH}
\mbox{P.N. Burrows\unskip,\iOXF}
\mbox{A. Calcaterra\unskip,\iFRAS}
\mbox{R. Cassell\unskip,\iSLAC}
\mbox{A. Chou\unskip,\iSLAC}
\mbox{H.O. Cohn\unskip,\iTENN}
\mbox{J.A. Coller\unskip,\iBU}
\mbox{M.R. Convery\unskip,\iSLAC}
\mbox{V. Cook\unskip,\iWASH}
\mbox{R.F. Cowan\unskip,\iMIT}
\mbox{G. Crawford\unskip,\iSLAC}
\mbox{C.J.S. Damerell\unskip,\iRAL}
\mbox{M. Daoudi\unskip,\iSLAC}
\mbox{S. Dasu\unskip,\iWISC}
\mbox{N. de Groot\unskip,\iBRI}
\mbox{R. de Sangro\unskip,\iFRAS}
\mbox{D.N. Dong\unskip,\iMIT}
\mbox{M. Doser\unskip,\iSLAC}
\mbox{R. Dubois\unskip,}
\mbox{I. Erofeeva\unskip,\iMOSCOW}
\mbox{V. Eschenburg\unskip,\iMISSI}
\mbox{E. Etzion\unskip,\iWISC}
\mbox{S. Fahey\unskip,\iCOLO}
\mbox{D. Falciai\unskip,\iFRAS}
\mbox{J.P. Fernandez\unskip,\iUCSC}
\mbox{K. Flood\unskip,\iMASS}
\mbox{R. Frey\unskip,\iOREG}
\mbox{E.L. Hart\unskip,\iTENN}
\mbox{K. Hasuko\unskip,\iTOHO}
\mbox{S.S. Hertzbach\unskip,\iMASS}
\mbox{M.E. Huffer\unskip,\iSLAC}
\mbox{X. Huynh\unskip,\iSLAC}
\mbox{M. Iwasaki\unskip,\iOREG}
\mbox{D.J. Jackson\unskip,\iRAL}
\mbox{P. Jacques\unskip,\iRUTG}
\mbox{J.A. Jaros\unskip,\iSLAC}
\mbox{Z.Y. Jiang\unskip,\iSLAC}
\mbox{A.S. Johnson\unskip,\iSLAC}
\mbox{J.R. Johnson\unskip,\iWISC}
\mbox{R. Kajikawa\unskip,\iNAGO}
\mbox{M. Kalelkar\unskip,\iRUTG}
\mbox{H.J. Kang\unskip,\iRUTG}
\mbox{R.R. Kofler\unskip,\iMASS}
\mbox{R.S. Kroeger\unskip,\iMISSI}
\mbox{M. Langston\unskip,\iOREG}
\mbox{D.W.G. Leith\unskip,\iSLAC}
\mbox{V. Lia\unskip,\iMIT}
\mbox{C. Lin\unskip,\iMASS}
\mbox{G. Mancinelli\unskip,\iRUTG}
\mbox{S. Manly\unskip,\iYALE}
\mbox{G. Mantovani\unskip,\iPERU}
\mbox{T.W. Markiewicz\unskip,\iSLAC}
\mbox{T. Maruyama\unskip,\iSLAC}
\mbox{A.K. McKemey\unskip,\iBRUN}
\mbox{R. Messner\unskip,\iSLAC}
\mbox{K.C. Moffeit\unskip,\iSLAC}
\mbox{T.B. Moore\unskip,\iYALE}
\mbox{M. Morii\unskip,\iSLAC}
\mbox{D. Muller\unskip,\iSLAC}
\mbox{V. Murzin\unskip,\iMOSCOW}
\mbox{S. Narita\unskip,\iTOHO}
\mbox{U. Nauenberg\unskip,\iCOLO}
\mbox{H. Neal\unskip,\iYALE}
\mbox{G. Nesom\unskip,\iOXF}
\mbox{N. Oishi\unskip,\iNAGO}
\mbox{D. Onoprienko\unskip,\iTENN}
\mbox{L.S. Osborne\unskip,\iMIT}
\mbox{R.S. Panvini\unskip,\iVAND}
\mbox{C.H. Park\unskip,\iSOONG}
\mbox{I. Peruzzi\unskip,\iFRAS}
\mbox{M. Piccolo\unskip,\iFRAS}
\mbox{L. Piemontese\unskip,\iFERR}
\mbox{R.J. Plano\unskip,\iRUTG}
\mbox{R. Prepost\unskip,\iWISC}
\mbox{C.Y. Prescott\unskip,\iSLAC}
\mbox{B.N. Ratcliff\unskip,\iSLAC}
\mbox{J. Reidy\unskip,\iMISSI}
\mbox{P.L. Reinertsen\unskip,\iUCSC}
\mbox{L.S. Rochester\unskip,\iSLAC}
\mbox{P.C. Rowson\unskip,\iSLAC}
\mbox{J.J. Russell\unskip,\iSLAC}
\mbox{O.H. Saxton\unskip,\iSLAC}
\mbox{T. Schalk\unskip,\iUCSC}
\mbox{B.A. Schumm\unskip,\iUCSC}
\mbox{J. Schwiening\unskip,\iSLAC}
\mbox{V.V. Serbo\unskip,\iSLAC}
\mbox{G. Shapiro\unskip,\iLBL}
\mbox{N.B. Sinev\unskip,\iOREG}
\mbox{J.A. Snyder\unskip,\iYALE}
\mbox{H. Staengle\unskip,\iCSU}
\mbox{A. Stahl\unskip,\iSLAC}
\mbox{P. Stamer\unskip,\iRUTG}
\mbox{H. Steiner\unskip,\iLBL}
\mbox{D. Su\unskip,\iSLAC}
\mbox{F. Suekane\unskip,\iTOHO}
\mbox{A. Sugiyama\unskip,\iNAGO}
\mbox{A. Suzuki\unskip,\iNAGO}
\mbox{M. Swartz\unskip,\iJHU}
\mbox{F.E. Taylor\unskip,\iMIT}
\mbox{J. Thom\unskip,\iSLAC}
\mbox{E. Torrence\unskip,\iMIT}
\mbox{T. Usher\unskip,\iSLAC}
\mbox{J. Va'vra\unskip,\iSLAC}
\mbox{R. Verdier\unskip,\iMIT}
\mbox{D.L. Wagner\unskip,\iCOLO}
\mbox{A.P. Waite\unskip,\iSLAC}
\mbox{S. Walston\unskip,\iOREG}
\mbox{A.W. Weidemann\unskip,\iTENN}
\mbox{E.R. Weiss\unskip,\iWASH}
\mbox{J.S. Whitaker\unskip,\iBU}
\mbox{S.H. Williams\unskip,\iSLAC}
\mbox{S. Willocq\unskip,\iMASS}
\mbox{R.J. Wilson\unskip,\iCSU}
\mbox{W.J. Wisniewski\unskip,\iSLAC}
\mbox{J.L. Wittlin\unskip,\iMASS}
\mbox{M. Woods\unskip,\iSLAC}
\mbox{T.R. Wright\unskip,\iWISC}
\mbox{R.K. Yamamoto\unskip,\iMIT}
\mbox{J. Yashima\unskip,\iTOHO}
\mbox{S.J. Yellin\unskip,\iUCSB}
\mbox{C.C. Young\unskip,\iSLAC}
\mbox{H. Yuta\unskip.\iAOMORI}

\it
  \vskip \baselineskip                   % \bigskip did not work
%  \centerline{(The SLD Collaboration)}   % include collaboration name
%  \vskip \baselineskip
  \baselineskip=.75\baselineskip   % shrink the interline spacing
\iAOMORI
  Aomori University, Aomori, 030 Japan, \break
\iBRI
  University of Bristol, Bristol, United Kingdom, \break
\iBRUN
  Brunel University, Uxbridge, Middlesex, UB8 3PH United Kingdom, \break
\iBU
  Boston University, Boston, Massachusetts 02215, \break
\iCOLO
  University of Colorado, Boulder, Colorado 80309, \break
\iCSU
  Colorado State University, Ft. Collins, Colorado 80523, \break
\iFERR
  INFN Sezione di Ferrara and Universita di Ferrara, I-44100 Ferrara, Italy,
\break
\iFRAS
  INFN Laboratori Nazionali di Frascati, I-00044 Frascati, Italy, \break
\iJHU
  Johns Hopkins University,  Baltimore, Maryland 21218-2686, \break
\iLBL
  Lawrence Berkeley Laboratory, University of California, Berkeley, California
94720, \break
\iMASS
  University of Massachusetts, Amherst, Massachusetts 01003, \break
\iMISSI
  University of Mississippi, University, Mississippi 38677, \break
\iMIT
  Massachusetts Institute of Technology, Cambridge, Massachusetts 02139, \break
\iMOSCOW
  Institute of Nuclear Physics, Moscow State University, 119899 Moscow, Russia,
\break
\iNAGO
  Nagoya University, Chikusa-ku, Nagoya, 464 Japan, \break
\iOREG
  University of Oregon, Eugene, Oregon 97403, \break
\iOXF
  Oxford University, Oxford, OX1 3RH, United Kingdom, \break
\iPERU
  INFN Sezione di Perugia and Universita di Perugia, I-06100 Perugia, Italy,
\break
\iRAL
  Rutherford Appleton Laboratory, Chilton, Didcot, Oxon OX11 0QX United Kingdom,
\break
\iRUTG
  Rutgers University, Piscataway, New Jersey 08855, \break
\iSLAC
  Stanford Linear Accelerator Center, Stanford University, Stanford, California
94309, \break
\iSOONG
  Soongsil University, Seoul, Korea 156-743, \break
\iTENN
  University of Tennessee, Knoxville, Tennessee 37996, \break
\iTOHO
  Tohoku University, Sendai, 980 Japan, \break
\iUCSB
  University of California at Santa Barbara, Santa Barbara, California 93106,
\break
\iUCSC
  University of California at Santa Cruz, Santa Cruz, California 95064, \break
\iVAND
  Vanderbilt University, Nashville,Tennessee 37235, \break
\iWASH
  University of Washington, Seattle, Washington 98105, \break
\iWISC
  University of Wisconsin, Madison,Wisconsin 53706, \break
\iYALE
  Yale University, New Haven, Connecticut 06511. \break

\rm
%
%  }   % end of address list

\end{center}

\newpage
% tables & figures follow here
%
\section*{$^*$ Tables and Figures}
\begin{table}[h]
\small
\begin{tabular}{|llcc|} \hline
Source & Parameter variation & $\delta A_b (\mu) $ & $\delta A_b (e)$  \\
\hline\hline 
Monte Carlo weights  & $f_b, f_c$ variation
                     &  $\pm$.006 & $\pm$.006 \\
Track efficiency     & MC-data multiplicity match
                     & $\pm$.008 & $\pm$.001 \\
Jet axis simulation  & 10 mrad smearing
                     & $\pm$.001 & $\pm$.010 \\
Background level     & $\pm 10\% (\mu), ~\pm 5\% (e)$
                     & $\pm$.003 & $\pm$.006  \\
Background asymmetry     & $ \pm 40\% $
                     & $\mp$.003 & $\mp$.004 \\
Neural net training  & 10 training runs 
		     & $\pm$.000 & $\pm$.012 \\
		     
BR($Z^0 \rightarrow b\bar{b}$)     & $R_b = .2173 \pm .0007$
                     & $\mp$.000 & $\mp$.000  \\
BR($Z^0 \rightarrow c\bar{c}$)     & $R_c = .1674 \pm .0038$
                     & $\pm$.001 & $\pm$.001  \\
BR($b \rightarrow l$)     & $(10.62 \pm 0.17)\% $
                     & $\mp$.004 & $\mp$.003  \\
BR($\bar{b} \rightarrow \bar{c} \rightarrow l$)     & $(8.07 \pm 0.25)\% $
                     & $\pm$.003 & $\pm$.003  \\
BR($b \rightarrow \bar{c} \rightarrow l$)     & $(1.6 \pm 0.4)\% $
                     & $\pm$.005 & $\pm$.001  \\
BR($b \rightarrow \tau \rightarrow l$)     & $(0.452 \pm 0.074)\% $
                     & $\pm$.002 & $<$.001  \\
BR($b \rightarrow J / \psi   \rightarrow l$)     & $(0.07 \pm 0.02)\% $
                     & $\pm$.003 & $\pm$.002  \\
BR($\bar{c} \rightarrow l$)     & $(9.85 \pm 0.32)\% $
                     & $\pm$.002 & $\pm$.002  \\
$B$ lept. spect. - $D^{**}$ fr.
                     & $(23 \pm 10)\%$, $B^{+}$,$B^{0}$;
                       $(32 \pm 10)\%$, $B_{s}$ 
                     & $\pm$.004 & $\pm$.003  \\
$D$ lept. spect.     & $ACCMM1 \: (^{+ACCMM2}_{-ACCMM3})$ \cite{syst} \
                     & $\pm$.004 & $\pm$.005  \\
$B$-tag		     & eff. calibration 
		     & $\pm$.014 & $\pm$.012 \\
L/D		     & DT/MC ratio 
		     & $\pm$.002 & $\pm$.000 \\ 

$B\rightarrow{D\bar D}$ 
		     &$(7.2 \pm 2.0)\%$
		     & $\pm$.008 & $\pm$.000 \\

$D^0/D^{\pm}$        &$15\%$ uncertainty
		     & $\pm$0.000 & $\pm$.001 \\

$B_{s}$ fraction in $b\bar{b}$ event     
                     & $.115 \pm .050$
                     & $\pm$.002 & $\pm$.005 \\
$\Lambda_{b}$ fraction in $b\bar{b}$ event     
                     & $.072 \pm .030$
                     & $\pm$.002 & $\pm$.003  \\
$b$, $c$ fragmentation   & $\epsilon_b=.0045$-$.0075$
                     & $\pm$.003 & $\pm$.002 \\
Aleph fragmentation   & $\epsilon_c=.045$-$.070$
                     & $\mp$.003 & $\mp$.003  \\
Polarization         & $<\!P_e\!> = .729 \pm .0038$ 
                     & $\mp$.005 & $\mp$.006  \\
Second order QCD     & $\Delta_{QCD}$ uncertainty
                     & $\pm$.005 & $\pm$.005  \\
gluon splitting      & $g_{b\bar b}$, $g_{c\bar c}$ uncertainty
		     & $\pm$.002 & $\pm$.003  \\

$B$ mixing $\chi$    & $\chi = .1186 \pm .0043 $
                     & $\pm$.015 & $\pm$.012  \\
$A_c$ 		     & $0.667 \pm 0.040$ 
		     & $\pm$.002 & $\pm$.005 \\

\hline %\tableline
Total Systematic   & & .026 & .028  \\
\hline
\end{tabular}
\caption{Systematic errors for the maximum likelihood analysis (1993-98)}
\label{systable}
\end{table}

\newpage
\vskip0.2in

\smallskip
\begin{figure}[hb!]
\parbox{2.7in} {
\epsfxsize=2.7in  
\epsfysize=2.7in
\epsfbox{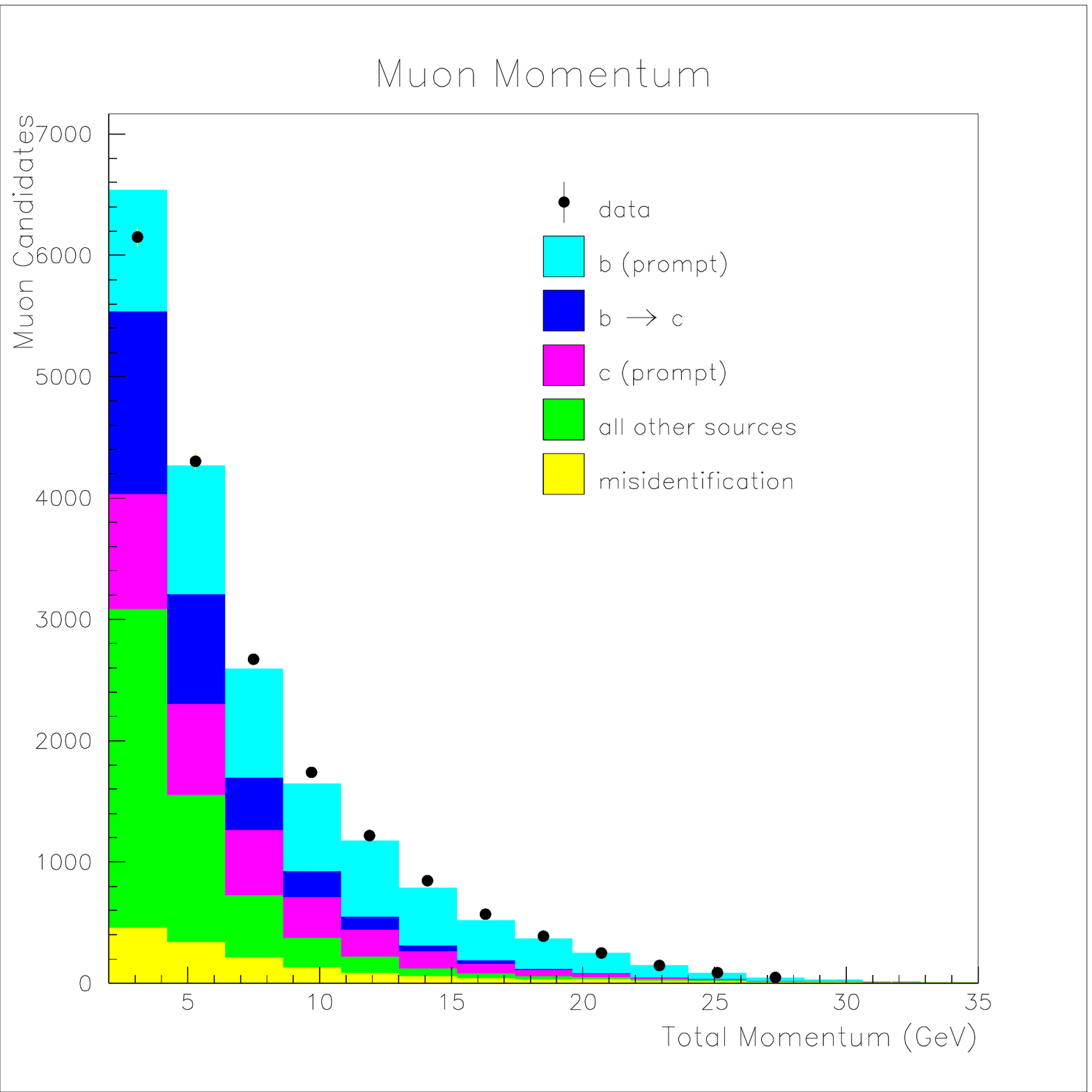}
\caption{Total momentum distribution of identified muons in 1996-98 data (dots) and Monte Carlo (histogram).}
\label{pmuon}}\hglue 0.3in\parbox{2.7in}{
\epsfxsize=2.7in   
\epsfysize=2.7in
\epsfbox{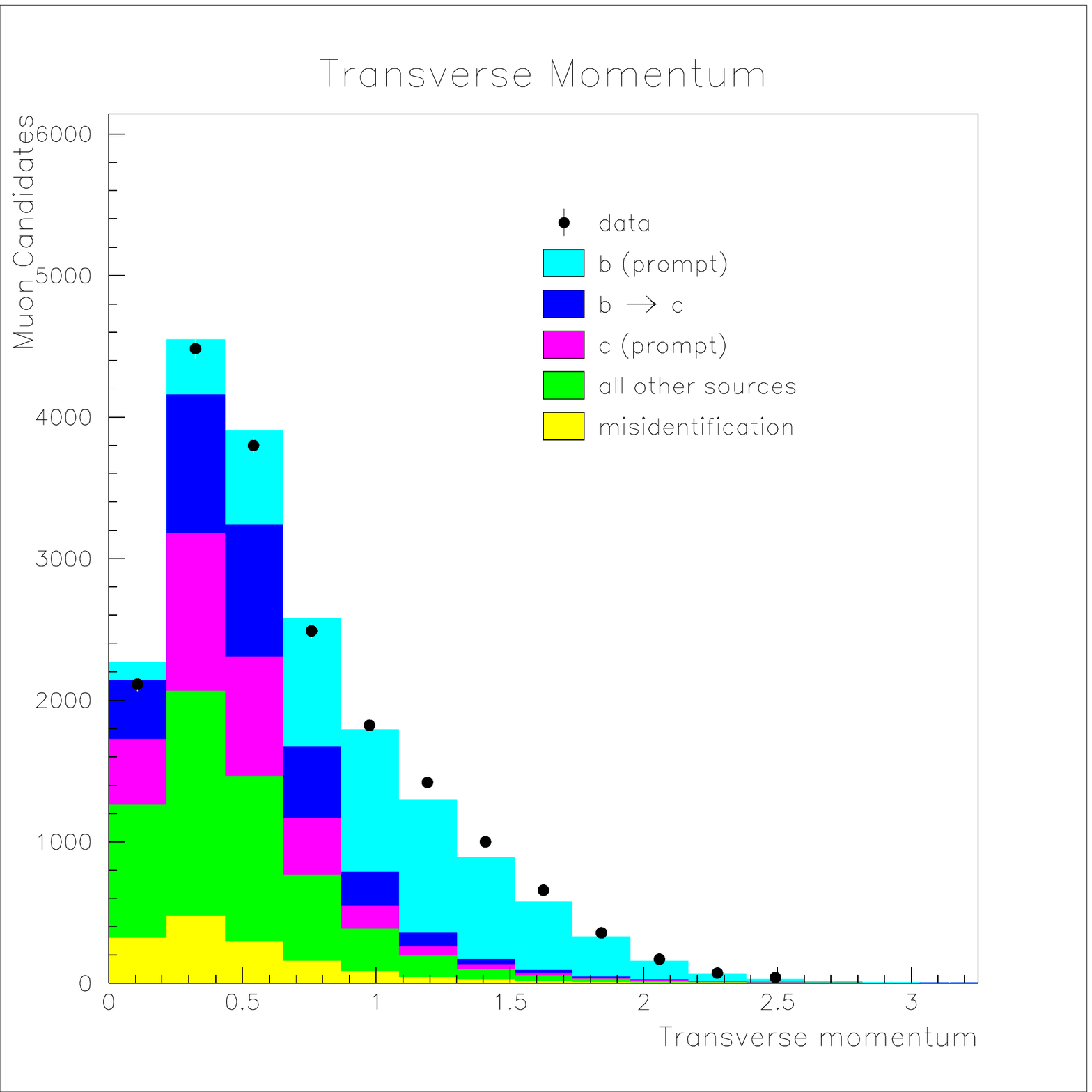}
\caption{Transverse momentum distribution of identified muons in 1996-98 data (dots) amd Monte Carlo (histogram).}
\label{Ptmuon}}
\end{figure}

\vfill \eject

\end{document}